\documentclass[prb,floatfix,twocolumn,amsmath,amssymb,showpacs]{revtex4}
\usepackage{graphicx}
\usepackage{bm}
\def\mbf#1{\mathchoice{\hbox{\boldmath $\displaystyle #1$}}
{\hbox{\boldmath $\textstyle #1$}}
{\hbox{\boldmath $\scriptstyle #1$}}
{\hbox{\boldmath $\scriptscriptstyle #1$}}}
\begin{document}

\title{
Spectral gap induced by structural corrugation in armchair
graphene nanoribbons 
}

\author{S. Costamagna, O. Hernandez and A. Dobry}
\affiliation{
Instituto de F\'{\i}sica Rosario, Consejo Nacional de
Investigaciones Cient\'{\i}ficas y T\'ecnicas,\\
Universidad Nacional de Rosario, Rosario, Argentina}

\date{\today}


\begin{abstract}

We study the effects of the structural corrugation or rippling on the electronic properties
of undoped armchair graphene nanoribbons (AGNR). 
First, reanalyzing the single corrugated graphene layer 
we find that the two inequivalent Dirac points (DP), 
move away one from the other.
Otherwise, the Fermi velocity $v_F$ decrease by increasing rippling. 
Regarding the AGNRs, whose metallic behavior depends on their width,
we analyze in particular the case of the zero gap band-structure AGNRs.  
By solving the Dirac equation with the adequate boundary condition we show that
due to the shifting of the DP  a gap opens in the spectra.
This gap scale with the square of the rate between 
the high and the wavelength of the deformation.
We confirm this prediction by exact numerical solution of the finite width rippled AGNR.
Moreover, we find that the quantum conductance, calculated by the 
non equilibrium Green's function technique vanish when the gap open. 
The main conclusion of our results is that a conductance gap should appear
 for all undoped corrugated AGNR independent of their width.       
\end{abstract}



\maketitle

\section{Introduction}
\label{intro}

The spectacular interest that have been raised  from the recent isolation
of an atomic Carbon layer, the graphene\cite{novoselov},
is based on the unusual dynamics the electrons have in this material.
In graphene, the electrons behaves as massless relativistic particles giving
rise, for example, to a sequence of Hall plateau quite different than the one observed
in 2D electrons systems confined in semiconductor heterojuntions. 
Also the transport properties are remarkable, the mobility of 
electrons in suspended graphene could be even higher than in 
any known semiconductor\cite{review-castro}.
Even the existing of graphene as a two-dimensional atomic crystal and its
stability under ambient conditions is an surprising fact. According to
the Mermin-Wagner theorem, there is not should be long-range crystalline order in
two dimensions at finite temperature. 
Even more, flexible membrane embedded in three-dimensional 
space should be crumpled because of long-wavelength bending fluctuations. 
However these fluctuations can be suppressed by anharmonic coupling between
bending and stretching modes. As a result, single-crystalline
membranes can exist but should be rippled\cite{reviewmembrane}.
Indeed, ripples were observed in graphene\cite{suspended}. 
It has been proposed that they should play an important 
role in its electronic properties\cite{Morozov,castro-1}.
In particular, intrinsic rippling has been proposed as one of the 
possible mechanism for electron scattering to explain the variation 
of the resistivity with the number of charge carrier 
experimentally seen in graphene\cite{Katsnelson}. 

Between the research areas of graphene, a very important one is the study 
of nanoribbons where the sheets are catted with a particular pattern 
to manage the electrical properties. 
Depending on the type of the border edges they can be either
in, namely, zigzag (ZGNR) or armchair configurations. 
It is known that a tight binding model predict that ZGNRs are always 
metallic while AGNRs can be either metallic or semiconductor, 
depending on their width\cite{Brey}.
However,  experiments suggest that AGNR are always insulators with a gap that scale 
with the inverse of its width\cite{Kim}. %
The effect of the corrugation on the electronic properties of graphene has been 
studied in previous works by a tight binding type model\cite{Guinea} and 
by ab-initio LDA calculation\cite{midgap}. 
The rippling induce a nonuniform 
gauge field whose effects has been analyzed by analogy with an applied magnetic field. 
In this sens a pseudo Landau levels (LL) were predicted. 
Regarding nanoribbons the effects of the rippling on the conductivity 
were analyzed in a model which include
also the effect of the disorder produced by charged impurities\cite{Klos}. 
Although this approach is quite realistic it 
can not isolate only the effect of the
rippling in order to known the contribution of each perturbation separately.
Therefore, in this work, our mainly purpose is to analyze the effect of the corrugation
on the electronic structure and the transport properties of AGNRs. 
First, as a necessary previous step we study 
the electronic properties of corrugated graphene layers.
We reanalyze the condition for the appearance 
of a flat band associated with the zero LL.
We show that an strong rippling on the sheet is necessary to produce such a flat band. 
Then, we focus on the armchair border type nanoribbons. 
Particularly we analyze the case of the zero gap band-structure AGNRs. 
We show that due to rippling an spectral gap is open in otherwise conducting ribbons.
We subsequent analyze the quantum conductance by the non equilibrium 
green's function technique\cite{dattabook} (NEGF) and show that the opening of 
the gap manifest in a insulating behavior of the undoped samples.    

The paper is organized as follows. In Section \ref{ripplesheet} we
provide a detailed explanation of the corrugation model adopted and
we study the effect of the corrugation
on the electronic spectra of a graphene sheet.
Then, in Section \ref{AGNR} we study
the effect of the rippling on the electronic spectra 
and the conduction properties of metallic AGNR. 
Finally, in Section \ref{conclusions} we present the conclusions of our paper
and we discuss it implicants.


\section{The effect of the rippling: the graphene sheet}
\label{ripplesheet}

Let us start by studying the effects of the corrugation on 
the electronic states for the case of 2D  graphene layer.
This question was examined in a series of previous papers\cite{Guinea,midgap}.
As it will be the starting point of our study of a corrugated nanoribbons
in the next Section, we reanalyze this problem in the present Section.
This will be presented after a detailed description of the model adopted.

\subsection{The model}
\label{model}

We describe the electronic properties of graphene layer
by means of a tight-binding model with one $\pi$-orbital for each Carbon atom 
and nearest-neighbor hopping between them.
Carbon atoms arrange in a honeycomb lattice.
It is not a Bravais lattice but can be constructed from the
hexagonal lattice by putting a basis of two atoms. 
The hexagonal lattice could be constructed as ${\bf R}^{\bf l}=l_x {\bf a}_1+l_y 
{\bf a}_2$ with $l_x$ and $l_y$ integers, ${\bf a}_1=\frac{a}{2}(3,\sqrt{3})$ and
${\bf a}_2=\frac{a}{2}(3,-\sqrt{3})$ are the primitive vectors.
The  basis is given by the atom A sited at ${\bf R}^{\bf l}$ and B at
${\bf R}^{\bf l}+\frac{a}{2}(1,\sqrt{3})$. {\it a} $\sim 1.41 \AA{}$ is the 
lattice constant and hereafter will be taken as unit of distances. 

\begin{figure}[hbt]
\vspace{0.7cm}
\includegraphics[width=0.44\textwidth]{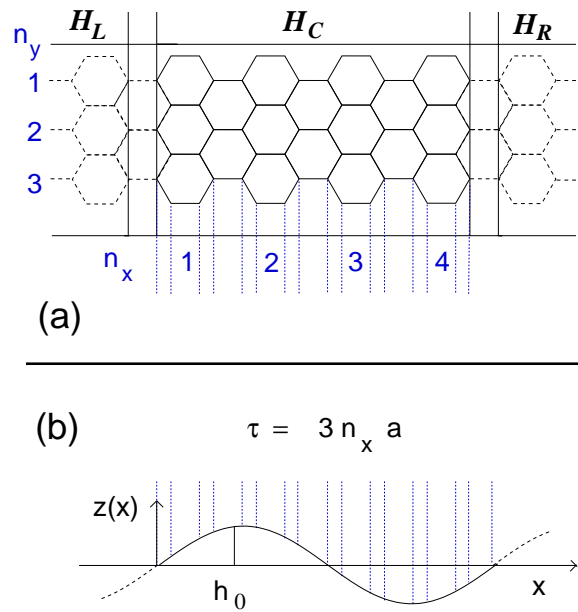}
\caption{(a) Schematic view of an armchair graphene nanoribbon.
Vertical lines separates the regions adopted for the conductance calculation
described in Sec.\ref{QCond} trough the Hamiltonians displayed on top. 
The central region possesses $n_x=4$ and $n_y=3$, listed as indicated.
(b) Side view in the $x-z$ plane of the displacement of Carbon atoms due to rippling. 
$h_0$ is the amplitude and $\tau$ the period. {\it{a}} is the lattice constant of graphene.
Vertical dotted lines from (a) to (b) 
gives reference for the actual positions of Carbon atoms
in the corrugated nanoribbon.
}
\label{fig1}
\end{figure}

The corrugation was included as a sinusoidal function which module
the $z$ coordinate of each lattice site. 
The $x$ and $y$ coordinates remain into its values
in the honeycomb lattice.
For simplicity, we assume that rippling only depend on the $x$ direction 
(see fig.  \ref{fig1} for the case of a nanoribbon).
Then, the adopted $z(x)$ function reads:
\begin{equation}
z(x)=h_0  sin\Bigg(\frac{2\pi}{\tau}  x\Bigg) 
\label{corrfun}
\end{equation} 

\noindent where $h_0$ is the amplitude and $\tau$ the period
of the rippling.
This simple model for the corrugation has been used in previous works\cite{midgap,Guinea}.
As a superposition of functions of the type (\ref{corrfun}) 
with different wave vectors and different amplitudes could generate a quite general corrugation,
our study should be taken as a starting point for the effect of a general corrugated function. 
Evermore as we point to the study of nanoribbons
which by definition are much longer than wider 
the one-dimensional character of the rippling function is justified.

The modulated structure induce an spacial variation of the hopping
parameter $t$, being now dependent on the distance between the atoms.
%
Developing $t$ up to first order in the perturbed distance
between the nearest neighbor ($\Delta a$)  we have:
\begin{eqnarray}
t &=& t_0 + \frac{t_0}{a}\alpha  \Delta a \nonumber\\
\Delta a&=& \sqrt{a+(z(x)-z(x'))^2}-a
\end{eqnarray} 

\noindent where $t_0\approx 2.66 eV$ is the  hopping parameter of the undeformed
graphene which is taken as a unit of energy  in the present work.
$\alpha=\frac{\partial \log t}{\partial \log a}\sim 2$, 
whose value is taken from Ref. (\onlinecite{review-castro}).
The rippling also produce a bending of the p-orbitals. However it has been shown that the 
relative change of the hopping due to the bending is weaker than the one due to the change of the bond length\cite{Klos}.
For simplicity, we neglect this effect in the present paper.  

The tight binding Hamiltonian in the presence of the corrugation becomes:

\begin{equation}
H  = - \sum_{\mbf{l},\mbf{\delta},\sigma} t[z_{B{\mbf l}+\mbf{\delta}}-z_{A{\mbf l}}]
 (c_{\mbf{l} A,\sigma}^\dagger c_{_{\mbf{l}+\mbf{\delta} B,\sigma}} + h. c.)
\label{hamtb}
\end{equation}
where $\mbf{l}=(l_x,l_y)$ denotes the integer coordinates in the hexagonal Bravais lattice 
and $\mbf{\delta}=\{(0,0),(-1,0),(-1,1)\}$.

\subsection{Dirac equation}
\label{dirac1}

A low energy Hamiltonian could be obtained from (\ref{hamtb}) expanding the Fourier transform
of the electron operators around two of the inequivalent points where the 
dispersion relation vanish, namely the Dirac points (DP).
We choose these points as given by
${\bf K}=\frac{2\pi}{3 a}(1,\frac{1}{\sqrt{3}})$ and 
${\bf K}'=\frac{2\pi}{3 a}(1,-\frac{1}{\sqrt{3}})$.
In addition we assume a smooth variation of $z(x)$, i.e. $\tau >> a$.
The  low energy physics is described by the sum of two Dirac
Hamiltonians for massless particles in a pseudo-magnetic field.
It is given by:
\begin{eqnarray}
H &=&  v_F \int dr^2 ( \mbf{\Psi}_{1}^\dagger (r) \mbf{\sigma}^* \cdot (-i\mbf{\nabla}+ \mbf{A}(r))
 \mbf{\Psi}_{1}(r)
 \nonumber\\&+&\mbf{\Psi}_{2}^\dagger (r) \mbf{\sigma} \cdot (-i \mbf{\nabla}- \mbf{A}(r))
 \mbf{\Psi}_{2}(r))
 \label{DiracH}
\end{eqnarray} 

\noindent where $v_F=\frac{3 t_0 a}{2 }$ is the Fermi velocity, 
$\mbf{\sigma}=(\sigma_x,\sigma_y)$ is the vector of the 
Pauli matrices and $\mbf{\sigma}*$ its complex conjugate.
The two component field is $\mbf{\Psi}_{i}=(c_{iA},c_{iB})$ 
with $c_{i,A(B)}$  creation field operator over the A(B) sublattice
The index $i=1,2$ refers to states with momentum near $\bf{K}$, $\bf{K}'$.
The gauge vector $\mbf{A}(r)=(A_x,A_y)$ were induced by the rippling and is given by:
\begin{eqnarray}
A_x&=&0 \nonumber\\
A_y&=&-\frac{\alpha}{4}(\partial_x h)^2=-\frac{\alpha}{4}
(h_0 q)^2  (cos
\Big(\frac{4\pi}{\tau}x\Big)+\frac12)\nonumber\\
&\equiv& \hat{A}_y+k_0
\end{eqnarray}

\noindent $k_0=-\frac{\alpha}{8}
(h_0 q)^2$ is the $k=0$ Fourier component of $A_y$ and $q=\frac{2\pi}{\tau}$ the wave vector of the rippling.
Note that as the time reversal symmetry is not broken by the rippling, 
this implies that the coupling of $\mbf{\Psi}_1$ 
with $\mbf{A}(r)$ have different sign
than the one of
$\mbf{\Psi}_2$.
This fact will be important in the study of the nanoribbons undertaken in the next Section.

The Dirac equation in a magnetic field corresponding to the one particle problem generated by
eq. (\ref{DiracH}) at $i=1$ could be exactly solved for the zero energy eigenstate. By following similar steps as in Ref. 
\onlinecite{diracmg}
we find the following two degenerated eigenfunctions:
\begin{eqnarray}
\Psi_{+0}(x,y)=\frac{N}{\sqrt{L_y L_x}} e^{-i k_0 y} \Big( \stackrel{e^{F(x)}}{0}\Big) \nonumber\\
\Psi_{-0}(x,y)=\frac{N}{\sqrt{L_y L_x}}e^{-i k_0 y} \Big( \stackrel{0}{e^{-F(x)}}\Big)
\label{zerost}
\end{eqnarray}

\noindent with $F(x)=\frac{\alpha}{8}h_0^2 q \sin(2 q x)$ 
and $L_{x,y}$ the system length in both direction.
 $N$ is a normalization constant given by  $N=[I_0(\frac{\alpha h_0^2 q}{4})]^{-\frac12}$ with $I_0$ the modified Bessel function of zero order.

The low energy eigenstates of (\ref{DiracH}) can be studied by degenerated perturbation 
theory over the zero energy states $\Psi_{+0}(x,y)$ and $\Psi_{-0}(x,y)$. 
The potential $A_y$ is periodic with period $\frac{\tau}{2}$. 
Therefore the solutions of (\ref{DiracH}) fulfill the Bloch theorem. 
They will have the form:
\begin{eqnarray}
\Psi_{n,\mbf{k}}(x,y)= e^{i  \mbf{k}\cdot\mbf{r}}  \psi_{n,k_x}(x) 
\label{Bloch}
\end{eqnarray} 

\noindent where n is a band index and $k_x$ belong to the Brillouin zone 
of the imposed periodic lattice i.e. $-\frac{2 \pi}{\tau} \leq k_x < \frac{2 \pi}{\tau}$.
 Otherwise there is not restrictions on the values of $k_y$.
Note that (\ref{zerost}) is of the form of (\ref{Bloch}) with a quasi-momenta of the zero energy eigenstates given by $\mbf{k_0}=(0,-k_0)$.
and function $\psi_{0,k_x}(x)$ given by:
\begin{eqnarray}
\psi_{0,\mbf{k}}(x)= \frac{N}{\sqrt{L_y L_x}}  
\Big( \begin{array}{c}
a_0 e^{F(x)}\\
b_0 e^{-F(x)}\\
\end{array}
\Big)
\label{Bloch0}
\end{eqnarray} 
$a_0$ and $b_0$ satisfying $\sqrt{a_0^2+b_0^2}=1$.
In (\ref{Bloch}) the Bloch function fulfill $\psi_{n,k_x}(x+\frac{\tau}{2})=\psi_{n,k_x}(x)$ and satisfy
$H(\mbf{k})\psi_{n,\mbf{k}}=E_{n,\mbf{k}}\psi_{n,\mbf{k}}$, with $H(\mbf{k})$ given by:
\begin{eqnarray}
H(\mbf{k})= H_0(\mbf{k})+H_{int}(\mbf{k})\nonumber\\
H_0(\mbf{k})=v_F \mbf{\sigma}^*\cdot(-i\mbf{\nabla}+\hat{\mbf{A}}(\mbf{r}))\nonumber\\
H_{int}(\mbf{k})=v_F \mbf{\sigma}^*\cdot(\mbf{k}+\mbf{k_0})
\label{Heff}
\end{eqnarray} 
As  $H_0(\mbf{k})\psi_{0,\mbf{k}}(x)=0$ we identify  $H_0(\mbf{k})$ as the unperturbed Hamiltonian.
The approximated near-zero energy eigenstates could be obtained by diagonalizing $H_{int}$ in the subspace of the degenerated zero energy states. 
$H_{int}$ proyected onto this subspace is a $2\times2$ matrix called $\mbf{h}$ and given by:
 \begin{eqnarray}
\mbf{h}&=&
\hat{v}_F
\Bigg( \begin{array}{cc}
0  &  k_x + i(k_y+k_0)\\
k_x - i(k_y+k_0) &  0 \\
\end{array}
\Bigg)=\nonumber\\
&&=\  \hat{v}_F \mbf{\sigma}^*\cdot(\mbf{k}+\mbf{k_0})
\label{Heffproy}
\end{eqnarray} 
$\hat{v}_F=N^2 v_F$ is a renormalized Fermi velocity.
The eigenvalues of (\ref{Heffproy}) are:
\begin{eqnarray}
E_1(k_x,k_y)=\pm \hat{v}_F\sqrt{k_x^2+(k_y+k_0)^2 }
\label{reldis1}
\end{eqnarray} 

\begin{figure}
\includegraphics[width=0.3\textwidth]{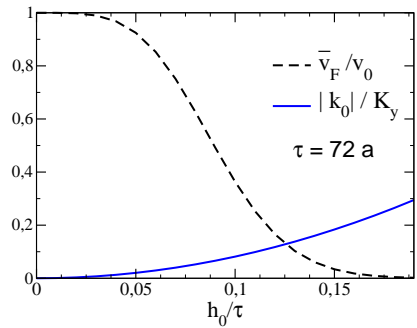}
\caption{Shifting of the DP $|k_0|$ normalized 
to the y coordinates of this point ($K_y$) as a function of $h_0/\tau$ (solid line). 
We also shown $\frac{\hat{v}_F}{v_F}$ giving the reduction of the Fermi velocity 
(dotted line).}
\label{vfk0}
\end{figure}

We see that the main effects of the corrugation are:
\begin{enumerate}
 \item DP sited in $\mbf{K}$ have been shifted to  $\mbf{K}-\mbf{k_0}$.
\item $\hat{v}_F$ decrease by increasing rippling.
\end{enumerate}
If we had studied the behavior near $\bf{K'}$ by analyzing the second 
term in (\ref{DiracH}) we would have obtained instead of eq. (\ref{reldis1})
the result: 
\begin{eqnarray}
E_2(k_x,k_y)=\pm \hat{v}_F\sqrt{k_x^2+(k_y-k_0)^2 }
\label{reldis2}
\end{eqnarray} 
Therefore $\mbf{K}'$ moves opposite than $\mbf{K}$. As $k_0$ is a negative quantity the two DP 
 away each other in presence of the rippling. 
In Fig. \ref{vfk0} we show the relationship $\frac{|k_0|}{K_y}$ and $\frac{\hat{v}_F}{v_F}$ as a function of $\frac{h_0}{\tau}$ 


\begin{figure}[hbt]
\vspace{0.6cm}
\includegraphics[width=0.45\textwidth]{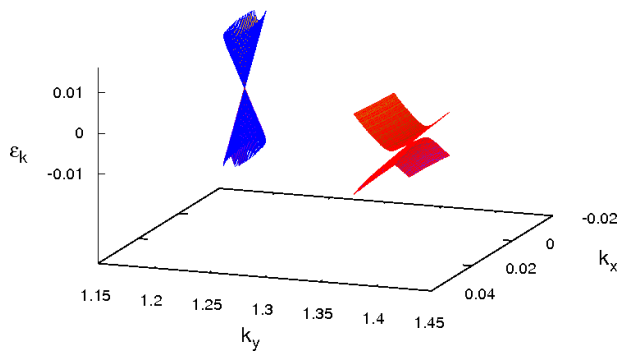}
\caption{Low energy band structure of uncorrugated (blue)
and corrugated (red) graphene layers.
The corrugation parameters are $\tau =$ 72 {\it a} and $h_0$ equals $0.0$, which correspond to
the Dirac cone, and $6.0$ in units of {\it{a}}, respectively.
This result qualitatively agree with the obtained by means the Dirac equation
including the pseudo-magnetic field.
}
\label{gf2}
\end{figure}

For completeness, let us provide a comparison, just at a qualitative level, of
the above findings with the band structure of corrugated graphene layers.
Since we are trying to detect the displacement of the Dirac points,
first we must compute the band structure of the uncorrugated case.
Note here that since the adopted ripple develops in the x axis direction with a
period equal to $\tau$, as can be seen in fig.\ref{fig1} (b),
then the unit cell must be extended up to include the whole rippled region
from where the hopping becomes repeated again.
This enlargement of the unit cell produces thus a reduction of the $k_x$ components
of the Dirac points $K$ and $K'$ due to band reflexion
at the border of the original first Brilloun zone.
As can be seen in the fig.\ref{gf2}, where we show the low energy spectrum
obtained for corrugations whose period is $\tau =$ 72 {\it a} 
and with amplitudes $h_0$ 0.0 and $6.0$ in units of {\it{a}},
the corrugation effectively shifts the Dirac point,
located originally at $K$, in the $k_y$ direction.
Also, in addition to this shift and in accord with the earlier predictions,
the observed small slope of the dispersion bands of corrugated graphene
implies a lower Fermi velocity of electrons near the Fermi level.
It should be mention here that this result is related with the appearance of Landau like levels (LL) 
induced by rippling and studied in related works\cite{Guinea}. Note  that from Fig. \ref{vfk0} we see that large values of $\frac{h_0}{\tau}$
 are required to obtain a total flattening of the band. This could be interpreted as that the predicted zero LL  appears only when the rippling is strong enough.
This conclusion is consistent with the results of Ref. \onlinecite{midgap}. In this work it was shown that as a result of the relaxation of the structure, the corrugation decrease and the 
flat band disappears from the spectra.  


\section{The effect of the rippling: armchair graphene nanoribbons}
\label{AGNR}
Let us now turn into the main subject of this work which is the study of the effects of corrugations on the spectra of graphene nanoribbons.
As we have mentioned 
in the Introduction, a tight binding based description of the electronic properties of the ribbons predict that, depending on their width and border edge type, they can be metallic, with zero band gap, or insulators\cite{Brey}. 
For example, for armchair edge (fig.\ref{fig1}) there are specific values of the width has 
for which metallic behavior is expected. Otherwise they should be insulators. 
%
In spite of that, the experiments shown that graphene
nanoribbons are insulators independent of their width\cite{Kim}. 
Though device fabrication process does not give atomically precise control on the type of edges, 
this seems to show that a pure tight binding model is not enough to describe the GNR as it is for the sheets.
Both, electronic correlation\cite{brey-1} and edge disorder\cite{castro-2, edge-defects}, 
have been proposed as possibles mechanisms that give rise to this behavior.  
In this Section we show that rippling could give an additional source for the electronic gap, at least for the armchair edge type.  

\subsection{Dirac equation with boundary conditions}

As for the infinite graphene sheet,
it is possible to study the low energy electronic structure of graphene nanoribbons 
by a Dirac like equation, but in this case adequate boundary conditions must be imposed\cite{Brey}.
For armchair edges, the valley states near the DPs, ${\bf K}$ and ${\bf K'}$, 
get admixture and metallic behaviors are obtained for certain widths.
We have described in the previous Section that the rippling produce a shift 
of the two inequivalent DPs. 
As we will shown below, to include only this shifting is 
enough to obtain a drastic change in the electronic behavior 
of AGNRs due to corrugations.
%

The wavefunction for sub-lattices A (B) are therefore given by:

\begin{eqnarray}
\Phi_{A(B)}(\mbf{r})&=& e^{i  (\mbf{K-k_0})\cdot\mbf{r}}  \Psi_{1,A(B)}(\mbf{r})\nonumber\\
&+& e^{i  (\mbf{K'+k_0})\cdot\mbf{r}}  \Psi_{2,A(B)}(\mbf{r})
\label{wfnr}
\end{eqnarray} 

\noindent where $\Psi_{1,A(B)}$ are the components of the 
spinor wave function for states near $\mbf{K}$ and $\Psi_{2,A(B)}$ the ones near $\mbf{K'}$.
For AGNR the wave function should vanish at $y=0$ and $y=L_y$.

If we neglect other effect than the shifting of the DP, the translational 
symmetry is preserved and implies that $\Psi_{1(2),A(B)}(\mbf{r})$ could be written as:

\begin{eqnarray}
\Psi_{1(2),A(B)}(\mbf{r})&=& e^{i  k_x x} \phi_{1(2),A(B)}(y)
\label{phinr}
\end{eqnarray} 

The solutions of the Dirac equation with the previous stated 
boundary conditions (BC) has now the form:

\begin{eqnarray}
\phi_{1,B}&=&  e^{i  k_n y} \nonumber\\
\phi_{2,B}&=&- e^{-i  k_n y}, 
\label{phi12nr}
\end{eqnarray} 

\noindent with the energies given by:

\begin{eqnarray}
\epsilon=\pm {v}_F\sqrt{k_x^2+k_n^2 }.
\label{reldisNR}
\end{eqnarray} 

The values of $k_n$ fulfills $\sin[(k_n+K_y-k_0)L_y]=0$ and are given by:

 \begin{eqnarray}
k_n=\frac{n\pi}{L_y}-K_y+k_0
\label{conkn}
\end{eqnarray} 

\noindent In absence of $k_0$, $k_n$ could vanish giving rise to 
a gapless spectra as is seen from Eq. (\ref{reldisNR}). 
This is the case  when the width of the ribbons is 
$L_y=3(n_y-1)\sqrt{3}a$ with $n_y$ integer.
However when $k_0$ is present the condition for gapless spectra 
could not be fulfilled in general. Even more for
the $L_y$ and $n_y$ giving a gapless spectra in absence of the rippling 
we now have a dispersion of the form  
$\epsilon=\pm {v}_F\sqrt{k_x^2+k_0^2 }$ and therefore a gap given by:

\begin{eqnarray}
\Delta=2 v_F k_0\sim \left(\frac{h_0}{\tau}\right)^2
\label{gapscaling}
\end{eqnarray}

\noindent Thus, from the previous consideration we expect a gap for 
AGNR due to rippling that scale as $\left(\frac{h_0}{\tau}\right)^2$. 
A more accurate treatment of the rippling will correct the constant 
multiplying the scaling law and possible produce correction 
to this scaling for large enough rippling values.

\begin{figure}[hbt]
\vspace{0.3cm}
\includegraphics[width=0.37\textwidth]{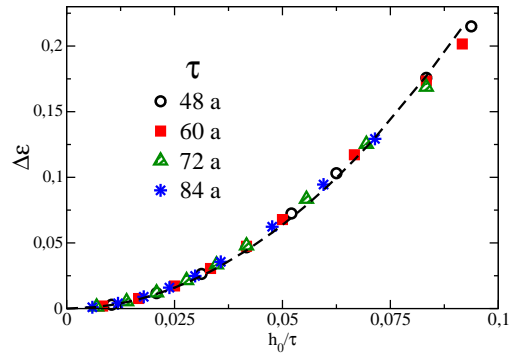}
\caption{Quadratic behaviour of the zero energy gap as a function of $h_0/\tau$ 
in corrugated armchair graphene nanoribbons. 
Values obtained from band structure calculations for the
different corrugation parameters sets displayed on the plot. 
The dotted-line is a fitting to a quadratic law (see text for details).}
\label{fig-gap}
\end{figure}

To confirm the previous predictions we have calculated numerically the eigenstates of 
corrugated armchair graphene nanoribbons. This was performed following the
steps described at the end of Sec. \ref{ripplesheet} taken into account 
here the armchair ribbon geometry displayed on fig. \ref{fig1}.
As we were interested in detecting the gap and in analyzing its
dependence with the corrugation,
we have explored a wide set of rippling parameters keeping fixed the width
of the ribbons in the values in which they possesses metallic behavior being uncorrugated.
Figure \ref{fig-gap} shows the zero energy gap as a function of $h_0/\tau$ for the various 
sets of parameters displayed on the plot, for a width given by $n_y =$ 8. Note that for the small 
values of $h_0/\tau$ included in this figure, the gap does not depende on $h_0$ and $\tau$ independently 
but on the ratio between these  quantities.
We have fitted the calculated gap using the least squares method for a function 
of the type $\Delta=C(\frac{h_0}{\tau})^2$ obtaining excellent agreement with correlation coefficients 
near one with a precision of $10^{-3}$. The obtained fitting is shown in Fig. \ref{fig-gap} with dotted line.
In accord to the stated above results, we have observed that by including higher values of $h_0/\tau$ 
the fitting become less 
precise. 

\subsection{Quantum Conductance}
\label{QCond}

Finally, we analyze how all the above findings affect
the electronic transport properties of the armchair graphene nanoribbons.
For this purpose, we employed the non-equilibrium Green's function 
formalism\cite{dattabook,brey-1} which provides an useful method to study
the quantum conductance in nanoscopic systems.

Before a discussion of the results let us present a briefly description of
the method and the implementation adopted. We considered one corrugated unit cell,
as central region $H_C$, connected to two semi-infinite outstanding
graphene nanoribbons flat leads, all having width length $n_y = 8$
which corresponds to a zero band gap structure in the uncorrugated case.
The hopping parameter in both leads Hamiltonians, $H_L$ and $H_R$,
being flat, is $t_0$. Hence, the whole system is then described by
\begin{equation}
H= H_C  + H_R + H_L + h_{LC} + h_{LR},
\end{equation}
\noindent where $h_{LC}$ and $h_{LR}$ are the hopping terms
from both leads to the central corrugated portion of nanoribbon, which we set equal to $t_0$
(see fig.\ref{fig1}(a)).
Within this formalism, the Landauer conductance $G(E)$ of the system,
in a zero bias approximation, is expressed as

\begin{equation}
G(E) = \frac{2e^2}{h}T(E)
\end{equation}
\noindent where $T(E)$ is the transmission function given by
$T(E)  = {\rm Tr} (\Gamma_L(E) \mathcal{G}_C (E)\Gamma_R(E)
\mathcal{G}_C^\dagger(E))$, with
$\Gamma_\ell=-2 \mathcal{I}m(\Sigma_\ell (E))$
$(\ell = L, R)$ being
the couplings of the corrugated graphene nanoribbon to the leads,
and $\mathcal{G}_C(E) = (E-H_C - \Sigma_L -\Sigma_R)^{-1}$
the total Green's function including the leads self energies,
$\Sigma_L$ and $\Sigma_R$. These self energies
must be calculated through the leads surfaces Green functions (SGF).
Although it was reported a very time efficient way to compute
the SGF of graphene nanoribbons leads (See Fig. 2 of Ref. \onlinecite{golizadeh}), 
for the sake of simplicity we implemented the recursive iteration procedure, given by
$g_\ell = (E -H_\ell-T_{\ell}^{\dagger} g_\ell T_\ell)^{-1}$,
where $H_\ell$ is the unit cell lead Hamiltonian and $T_\ell$
is the interlayer coupling in the semi-infinite lead.
For the small width nanoribbons studied here we have checked that a fast convergence
is obtained.
In the iterative procedure, we set the tolerance as $10^{-6}$.

%
%

\begin{figure}[hbt]
\includegraphics[width=0.32\textwidth]{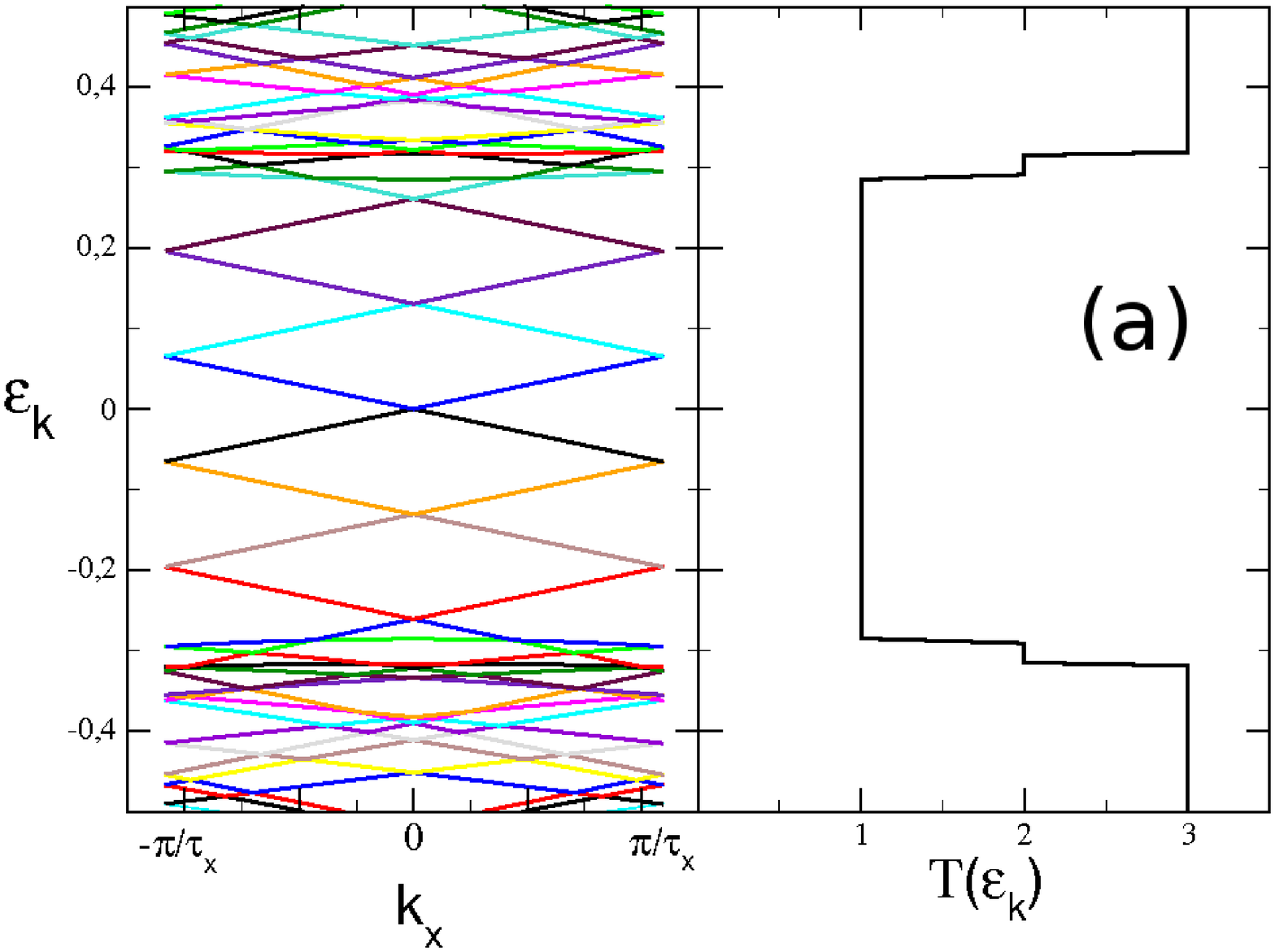}
\vspace{0.61cm}

\includegraphics[width=0.32\textwidth]{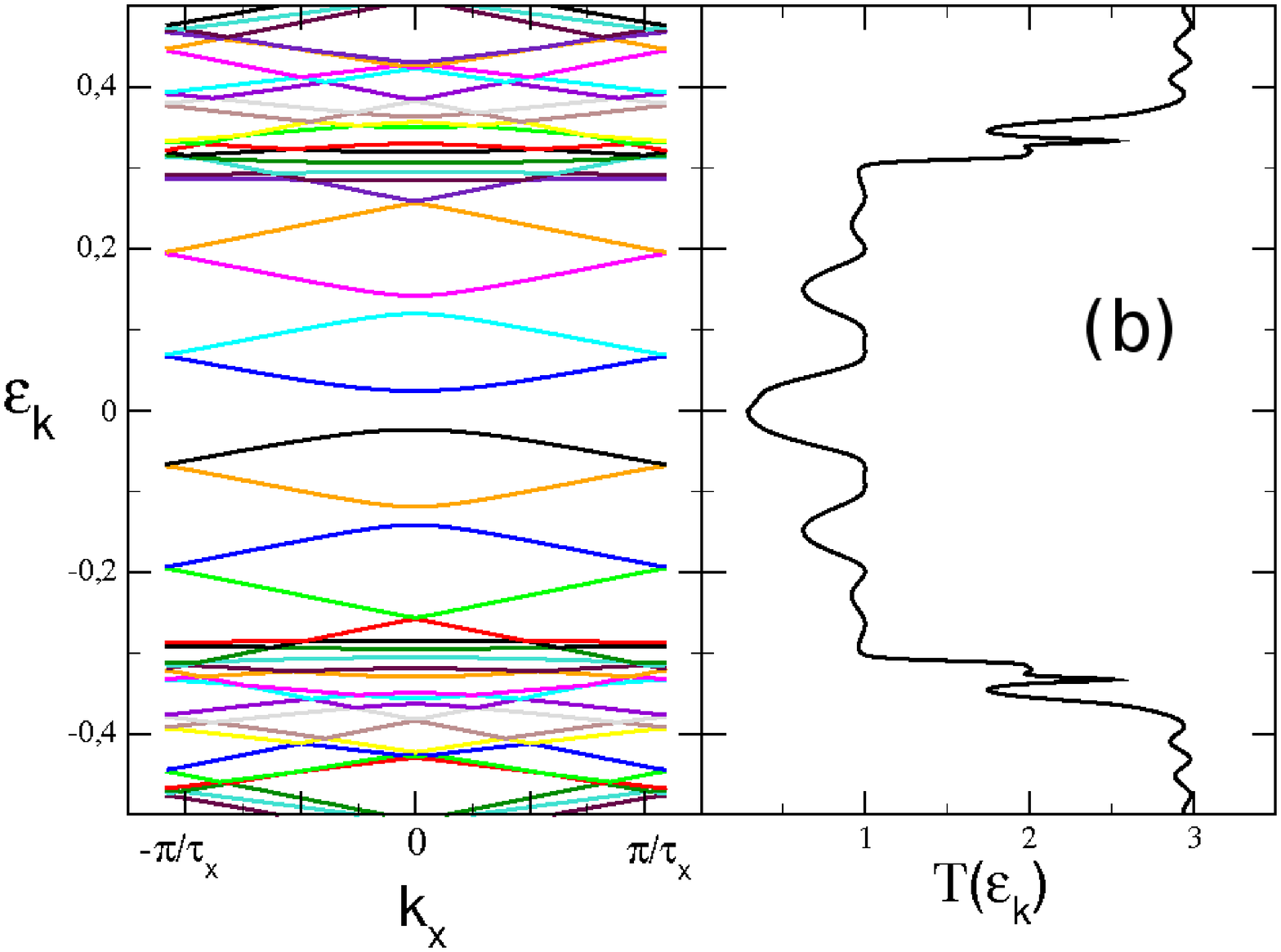}
\vspace{0.61cm}

\includegraphics[width=0.32\textwidth]{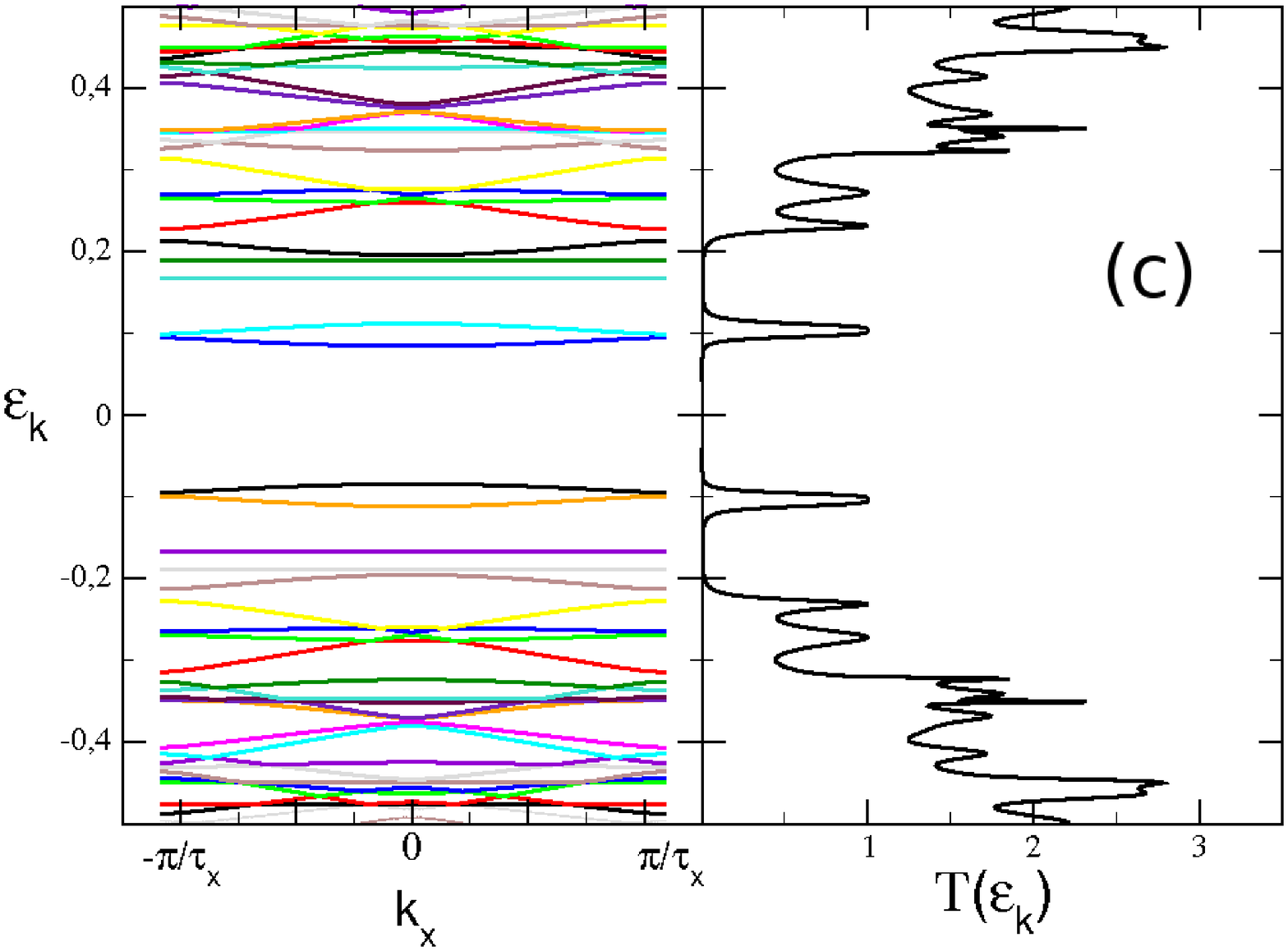}
\caption{Transmission coefficient of corrugated armchair graphene nanoribbons.
$n_x=24$, $n_y=8$ and $h_0=0$ (a), $h_0=3$ (b)
and $h_0=6$ (c), in units of {\it{a}} .
In each left panel we included the corresponding low energy band structure
which serve to a better analysis of the behavior of $T(E)$.
}
\label{fig5}
\end{figure}

Figures \ref{fig5} (a), (b) and (c), show the transmission
coefficients obtained for armchair border type nanoribbons whose
period $\tau$ is 72 {\it{a}} and with amplitudes $h_0=0.0, 3.0$ and $6.0$ in units of {\it{a}}, 
respectively.
In the left panels of the plots we included the corresponding low energy
bands structure computed numerically following the description given
at the end of Section \ref{ripplesheet} suited for the nanoribbon case.
The band reflexion at the first Brillouin zone can clearly be seen in
fig. \ref{fig5} (a) which correspond to the uncorrugated case.
Here, and such as is expectable, the transmission coefficient is equals to one at the Fermi energy.
In the fig. \ref{fig5} (b) the band gap at the Fermi energy appears,
and accordingly to this the transmission becomes reduced.
The fact that it do not reach zero is produced because the central region
possesses only one corrugated unit cell. Increasing the amplitude of the rippling
the gap becomes broader as is observed in fig. \ref{fig5} (c) and
the transmission goes to zero around the Fermi energy.
These results are consistent with the obtained above by
solving the Dirac equation.
Here also, a second null conductance region appears above a thin
perfect conductance interval at higher energies.
In addition to these results, obtained keeping fixed $\tau_x$ and varying $h_0$,
we calculated $T(E)$ for a wide range of corrugation periods and amplitudes,
finding in all cases the appearance of a gap at low energies around $E_F$.


\vspace{0.5cm}
\section{Conclusions}
\label{conclusions}

After having analyzed the Dirac equation including the pseudo-magnetic
field induced by the corrugation,
and having explored numerically a wide realistic range of rippling parameters sets,
we have found that the corrugations produces a gap in the low energy 
electronic spectrum of otherwise conducting armchair graphene nanoribbons.
We have found that this gap scale quadratic  with the rate between
the high and the wavelength of the deformation.
Accordingly to this, the quantum conductance $G(E)$ of the ribbons,
calculated by the NEGF formalism, vanish around the 
Fermi energy of undoped corrugated samples.

These results was preceded by an analysis of the corrugation effects over 
the electronic spectra of graphene layers.
Therein, we have found that
the corrugation just shift the Dirac points and renormalizes the Fermi velocity.
These findings agree with the results of numerical calculations.

In view of this, we conclude that the corrugations
are an important source of gap in the spectra in addition to the electronic 
correlations to which typically is assigned this effect\cite{brey-1}.
Although our study was performed modeling the corrugations by single 
sinusoidal functions, since any lattice deformation could be
approximated quite well by an adequate sum of sins, we strongly believe
that the results obtained here are extensible to general corrugations.
Moreover, our findings could be revealed in future experiments 
undertaken on the suspended clean sampled\cite{Bolotin} 
where extrinsic effect would be minimized. 
For these reasons we expect that our paper  motivates 
experimental works to determine 
the dependence of the gap  with the corrugation parameters that can be varied by changing the applied 
stress\cite{strains-suspended,uniaxial-strain} or the temperature.


\acknowledgments
This work was supported in part by grant 
PICT 1647 (ANPCYT).



\end{document}